\documentclass[12pt, onecolumn,draftcls]{IEEEtran}

\usepackage{cite}
\usepackage{graphicx}
\usepackage{amssymb,amsmath}
\usepackage{amsfonts}
\usepackage{multirow}
\usepackage{mathrsfs}
\usepackage{color}

\usepackage{algorithm}
\usepackage{algorithmic}
\usepackage{multirow}
\usepackage{amsmath}

\usepackage{graphics}
\usepackage{graphicx}
\usepackage{epsfig}
\usepackage{cases}
\usepackage{amsmath,bm}
\usepackage{wasysym}
\usepackage{subfigure}

\allowdisplaybreaks

\title{Distributed Clock Skew and Offset Estimation in Wireless Sensor Networks:
Asynchronous Algorithm and Convergence Analysis}
\author{Jian Du and Yik-Chung Wu
\thanks{Part of this manuscript appeared at the 2013 IEEE International Conference on
Acoustics, Speech, and Signal Processing \cite{JD13ICASSP}.
The authors are with the Department of Electrical and Electronic Engineering, The University of Hong Kong, Pokfulam Road, Hong Kong (e-mail: dujianeee@gmail.com, ycwu@eee.hku.hk).}}

\begin{document}
\newcommand*{\QEDA}{\hfill\ensuremath{\blacksquare}}  
\def\I{{\mathcal{I}}}
\def\arrow{{\rightarrow}}
\def\N{{\mathcal{N}}}
\def\B{{\mathcal{B}}}
\def\I{{\mathcal{I}}}
\def\F{{\bm F}} 

%
\maketitle
\begin{abstract}
In this paper, we propose a fully distributed algorithm for joint clock skew and offset estimation in wireless sensor networks based on belief propagation.
{In the proposed algorithm, each node can estimate its clock skew and offset in a completely distributed and asynchronous way: some nodes may update their estimates more frequently than others using outdated message from neighboring nodes.
In addition, the proposed algorithm is robust to random packet loss.}
Such algorithm does not require any centralized information processing or coordination,
and is scalable with network size.
{The proposed algorithm represents a unified framework that encompasses both classes of synchronous and asynchronous algorithms for network-wide clock synchronization.}
{It is shown analytically that the proposed asynchronous algorithm
converges to the optimal estimates with
estimation mean-square-error at each node approaching the centralized  Cram\'{e}r-Rao bound
 under any network topology.}
Simulation results further show that {the convergence speed is faster than that corresponding to a synchronous algorithm}.
\end{abstract}
\begin{keywords}
Clock synchronization, wireless sensor network, factor graph, asynchronous algorithm.
\end{keywords}
\newpage
\section{Introduction}\label{Section 1}
Wireless sensor networks (WSNs) have been widely used in environmental and emergency monitoring \cite{SensorSurvey, GBG},  event detection \cite{SmartGrid} and object tracking \cite{YCWuSP11}.
To perform distributed information processing in WSNs, a common clock across the network is usually required
to guarantee the nodes act in a collaborative and synchronized fashion.
Unfortunately, clock oscillator in each sensor node has its own imperfection and both  clock skew (frequency difference) and clock offset (phase difference) are present.
Therefore, time synchronization \cite{Swami} appears as one of the most important research challenges in the design of
WSNs.

Existing time synchronization algorithms can be categorized into two main classes.
One is pairwise synchronization {\cite{Kim11, Skog, LengMei11, AhmadIT, AhmadConf,Jeske05, Noh07, KYCheng,p2p1,p2p2,p2p3,netwide2}} where protocols are primarily designed to
synchronize two nodes.
The other is network-wide synchronization where protocols are designed to synchronize a large number of nodes in the network \cite{Ganeriwal,Noh08, pulse1,pulse2,pulse3,Consensus1, Consensus2, Consensus-WeihuaZhuang, Consensus3,Consensus-CDC12, LengMei11BP,netwide1,JD13ICASSP}.
Network-wide clock synchronization is much more challenging
due to limited radio range.
Nodes in a sensor network cannot directly communicate  with every other node,
but they have to do it via multi-hop.
Traditionally, network-wide clock synchronization in WSNs relies on spanning tree or clustered-based structure.
Under such structures, synchronization is achieved through layer-by-layer pairwise synchronization.
Such protocols, like time-synchronization protocol for sensor network (TPSN) \cite{Ganeriwal} and pairwise broadcast synchronization (PBS) \cite{Noh08},  suffer large overhead in building and maintaining the tree or cluster structure, and  are vulnerable to sudden node failures.

Without global structure or special nodes, by exchanging pulses emitted by oscillators, sensors are synchronized to transmit and receive at the same time in \cite{pulse1, pulse2, pulse3}.
However, these algorithms cannot provide a precise clock reading at the sensor node.
On the other hand, fully distributed synchronization based on averaged consensus algorithms have been proposed in \cite{Consensus1, Consensus2,Consensus-WeihuaZhuang, Consensus3,Consensus-CDC12,netwide1}.
Unfortunately, as shown in \cite{Consensus3, LengMei11BP}, consensus protocol is not optimal and the performance will deteriorate when message delay exists.
Besides, as average-consensus based algorithm seeks to reach  global average in the whole network, it has slow convergence \cite{Consensus-CDC12} (in order of hundreds of iterations before convergence).
More recently, \cite{LengMei11BP} pioneered the fully distributed network-wide clock offset estimation algorithm based on belief propagation (BP), and found that its performance is superior to consensus algorithms.
However, ignoring the effect of clock skew would significantly increase the re-synchronization frequency.
Moreover, {\cite{LengMei11BP} considers a parallel implementation with message exchange carried out in a synchronous
fashion}.
Notwithstanding, in many practical scenarios, the inter-sensor
message exchange is asynchronous since {random data packet losses} may occur,
and different nodes may update at different frequencies.
{At present, it is not clear the impact of these disturbance factors on the performance of synchronization algorithms.}

This work advances the state-of-the-art distributed synchronization in the following ways:
1) The distributed algorithm is fairly general and can cope with both clock skews as well as offsets over the whole network in parallel.
2) It represents a unified framework that encompasses both classes of synchronous \cite{LengMei11BP,JD13ICASSP} and asynchronous algorithms.
3) The convergence of the proposed method under asynchronous environments is formally proved.
The convergence result is derived for vector variable case, in which the Perron-Frobenious theorem used in \cite{LengMei11BP} is not applicable.
4) With the adoption of a different message passing rule from [29],
the mean-square error (MSE) performance of the derived algorithm is shown to approach the centralized Cram\'{e}r-Rao bound (CRB) asymptotically.  Simulations show that the convergence speed of asynchronous algorithm is faster than its synchronous counterpart.

The rest of this paper is organized as follows.
{The} system model is presented in Section \ref{Section 2}.
{A} fully distributed asynchronous clock skew and offset estimation algorithm based on BP is derived in Section \ref{Section 3}.
The convergence of the proposed asynchronous algorithm is analyzed in Section  \ref{Section 4}.
Simulation results are given in Section \ref{Section 5} and, finally, conclusions are drawn in Section \ref{Section 6}.

\textit{Notations}: Boldface uppercase and lowercase letters are used for matrices and vectors, respectively.
Superscript $T$ denotes transpose.
The symbol $\bm I_N$ represents the $N\times N$ identity matrix.
Notation $\N(\bm x|\bm \mu, \bm R)$ stands for the probability density function (pdf) of a Gaussian random vector $\bm x$ with mean $\bm \mu$ and covariance matrix $\bm R$.
The symbol $ \propto$ represents the linear scalar relationship between two real valued functions and
$|\mathcal{V}|$ denotes the cardinality of set $\mathcal{V}$.
{For two matrices $\bm X$ and $\bm Y$, $\bm X \succ \bm Y$ means that $\bm X - \bm Y$ is a positive definite matrix,
and $\bm X \succeq \bm Y$ means that $\bm X - \bm Y$ is a positive semi-definite matrix.}

\section{System Model}\label{Section 2}
Consider a general multi-hop sensor network with $M$ sensor nodes distributed in a field as shown in Fig. \ref{randomWSN}.
Let $\mathcal{V}=\{1,\ldots, M\}$ denotes the set of nodes and
$\mathcal{E}\subseteq \mathcal{V}\times \mathcal{V}$ is the set of edges.
An edge is denoted by $\{i, j\}$ if node $i$ and node $j$ can communicate directly.
In the example shown in Fig. \ref{randomWSN}, the vertices are depicted by circles and the edges by lines connecting these circles.
The set of neighbors of node $i$ is denoted by $\mathcal{I}(i)\subseteq \mathcal{V}$ with the definition that
$\mathcal{I}(i)\triangleq \{j\in \mathcal{V}| \{i,j\}\in \mathcal{E}\}$.
It is assumed that the radio coverage area of a node is circular with a specific radius so that each pair {of nodes} can exchange message only when their distance is less than both of their communication radiuses.
Furthermore, it is assumed that
any two distinct nodes can communicate with each other {through
a finite number of hops.
Such a network will be referred to as a strongly connected network.}

With {the} imperfection of oscillators and {possible} environmental changes, each node has a local clock with possibly different clock skew and offset. The relationship between real time $t$ and the local clock reading is modeled as
\begin{equation} \label{sec-order-model}
c_i(t) = \alpha_i t + \theta_i, \qquad i=1,\cdots, M,
\end{equation}
where $\alpha_i$ and $\theta_i$ are the clock skew and offset of node $i$, respectively.

To estimate and compensate such clock skews and offsets, {a} two-way time-stamp message exchange mechanism {was} proposed for pairwise clock synchronization \cite{Noh08}.
Specifically, as shown in Fig. \ref{TwoWay}, between one-hop neighboring nodes $i$ and $j$, at the $n^{th}$ round of time-stamp exchange, node $i$ sends a synchronization message to node $j$ at $t_n^1$ with its local clock reading $c_i(t^1_n)$ embedded in the message.
Node $j$ records its time $c_j(t^2_n)$ at the reception of that message and replies to node $i$ at $c_j(t^3_n)$.
The replied message contains both time stamps $c_j(t^2_n)$ and $c_j(t^3_n)$.
Then, node $i$ records the reception time from node $j$'s reply as $c_i(t^4_n)$.
$N$ rounds of such message exchange are performed between each pair of nodes to establish a relationship between the nodes $i$'s
and $j$'s clocks. In particular, for the $n^{th}$ round time-stamp exchange, we can write
\begin{equation} \label{twoequ1}
\frac{1}{\alpha_j}[c_j(t^2_n)-\theta_j] = \frac{1}{\alpha_i}[c_i(t^1_n)-\theta_i] + d_{i,j} + w_{j,n},
\end{equation}
and
\begin{equation} \label{twoequ2}
\frac{1}{\alpha_j}[c_j(t^3_n)-\theta_j] = \frac{1}{\alpha_i}[c_i(t^4_n)-\theta_i] - d_{j,i} - w_{i,n},
\end{equation}
where $ w_{j,n}$ and $w_{i,n}$ denote independent and identically distributed (i.i.d.) Gaussian random delay during the $n^{th}$ round of time-stamp exchange, with zero mean and variances $\sigma^2_j$, $\sigma^2_i$, respectively;
$d_{i,j}$ and $d_{j,i}$ represent the fixed message delay during {which} node $i/j$ sends message to node $j/i$, respectively.
Under the assumption that the network topology does not change during {the} clock synchronization process, we have $d_{i,j}=d_{j,i}$.
Adding  (\ref{twoequ1}) and (\ref{twoequ2}) and stacking all resultant equations for $N$ rounds of time-stamp exchange, we obtain
\begin{equation} \label{one-pair}
\bm A_{j,i}\bm \beta_j + \bm A_{i,j}\bm \beta_i = \bm z_{j,i},
\end{equation}
where $\bm A_{j,i}$ and $\bm A_{i,j}$ are $N$-by-$2$ matrices with the $n^{th}$ row being
$[c_j(t^2_n)+c_j(t^3_n), -2]$ and $-[c_i(t^1_n)+c_i(t^4_n), -2]$, respectively;
$\bm\beta_j\triangleq[\frac{1}{\alpha_j}, \frac{\theta_j}{\alpha_j}]^T$ and $\bm \beta_i\triangleq[\frac{1}{\alpha_i} , \frac{\theta_i}{\alpha_i}]^T$;
and $ \bm z_{j,i}$ is an $N$ dimensional vector with the $n^{th}$ element being $w_{j,n}-w_{i,n}$.
Since $w_{j,n}$ and $w_{i,n}$ are both i.i.d. Gaussian, it is easy to obtain $\bm z_{j,i}\sim \N(\bm z_{j,i}| \bm 0, \sigma_{i,j}^2\bm I_N)$, where $\sigma_{i,j}^2 = \sigma^2_i+\sigma^2_j$.
The goal is to establish global synchronization (i.e., estimate $\alpha_i$ and $\theta_i$ in each node) based on {the} local observations
$\bm A_{j,i}$ and $\bm A_{i,j}$.

\section{Asynchronous Distributed Estimation}\label{Section 3}
In this section, {the} asynchronous distributed clock parameter estimation algorithm is derived based on BP.
In the following, message exchange means BP message passing since two-way time-stamp exchange has been completed.
\subsection{BP Framework}
For the reason that the established clock relationships during two-way time-stamp exchanges involve interaction between neighboring nodes, the optimal clock estimate at each node requires the marginalization of joint posterior distribution of
all $\bm \beta_i$, which is
\begin{eqnarray}\label{marginal}
g_i(\bm \beta_i)\propto   \int...\int
\prod_{i=1}^M p(\bm\beta_i)
\prod_{{\{i, j\}\in \mathcal{E}}}
 p(\bm A_{i, j},\bm A_{j, i}|\bm \beta_i, \bm \beta_j)
d\bm \beta_1...d\bm \beta_{i-1}d\bm \beta_{i+1}d\bm \beta_{M},
\end{eqnarray}
where $p(\bm\beta_i)$ is the prior distribution of $\bm\beta_i$;
$p(\bm A_{i, j},\bm A_{j, i}|\bm \beta_i, \bm \beta_j)=\mathcal{N}(\bm A_{j, i}\bm\beta_j|\bm A_{i, j}\bm\beta_i, \sigma_{i,j}^2\bm I_N)$ is the likelihood function obtained from (\ref{one-pair}).
Node $1$ is assumed to be the reference node with
$p(\bm \beta_1)=\delta(\bm \beta_1-[1, 0]^T)$, and
its parameters need not to be estimated.
The computation of $g_i(\bm \beta_i)$ in (\ref{marginal})
needs to gather all information
in a central processing unit.
Besides, for the arbitrary network topology, the corresponding $|\mathcal{V}|$ and $|\mathcal{E}|$ can be very large leading to the {computationally demanding integration (\ref{marginal}).}

Although the joint posterior distribution of $\bm \beta_1,\ldots, \bm \beta_M$  (integrand in (\ref{marginal}))
is complicated due to {the } local interactions of sensor nodes, it is a product of local likelihood functions, each of
which depends on a subset of the variables.
Such a nice property
can be conveniently revealed in a factor graph \cite{Kschischang},
over which the computation of $g_i(\bm \beta_i)$
for all $i$ can be efficiently accomplished in a distributed way.
One example of factor graph is shown in Fig. \ref{FG}.
In this factor graph, local synchronization parameters $\bm \beta_i$, $i=1,\cdots, M$, are represented by variables nodes (circles).
If two sensor nodes $i$ and $j$ are within {the} communication range of each other, the corresponding {variables} $\bm \beta_i$ and
$\bm \beta_j$ are linked by a factor node (local function) $f_{i,j}=f_{i,j}\triangleq p(\bm A_{i, j},\bm A_{j, i}|\bm \beta_i, \bm \beta_j)$.
On the other hand, the factor node $f_{i}\triangleq p(\bm \beta_i)$ denotes the prior information.

The message passing algorithm operated on the factor graph involves two kinds of messages:
One is the message from factor node $f_{j,i}$ to a variable node $\bm \beta_i$,
defined as  \cite{Kschischang}
\begin{equation}
m^{(l)}_{f_{j,i} \arrow i}(\bf \beta_i)   \label{BPf2vs}
=  \int
m^{(l)}_{ j \arrow f_{j,i}}(\bf \beta_j)
f_{j,i}
d\bm \beta_j,
\end{equation}
where $l$ denotes the time of message exchange and $m^{(l)}_{ j \arrow f_{j,i}}(\bm \beta_j)$ is the other kind of message from {the} variable node to {the} factor node,
which is simply the product of the incoming messages on {the} other links, i.e.,
\begin{equation}
m^{(l)}_{j \arrow f_{j,i}}(\bm \beta_j) \label{BPvs2f}
 = \!\!\!\!\prod_{f\in\B(\bm \beta_j)\setminus f_{j,i}} m^{(l-1)}_{ f \arrow j}
 (\bm \beta_j),
\end{equation}
where $\B(\bm \beta_j)$  denotes the set of neighboring factors of $\bm \beta_j$ on the factor graph.
In particular, under such message computation rule,
the message from factor node $f_i$ to $\bm \beta_i$ {is} always equals to the prior distribution $p(\bm \beta_i)$ \cite{Kschischang}.

During the first round of message passing, it is reasonable to set initial messages from factor node to variable node
$m^{(0)}_{f_{i} \arrow  i}(\bm \beta_i)$ and $m^{(0)}_{f_{j,i} \arrow  i}(\bm \beta_i)$ as $p(\bm\beta_i)$ and non-informative message $\N(\bm\beta_i|\bm 0, +\infty\bm I_2)$, respectively.
Assuming $p(\bm \beta_i)=m^{(1)}_{f_{j,i} \arrow  i}(\bm \beta_i)$ is in Gaussian form (if there is no prior information, we can set the mean to be zero and set the variance to be a large value, i.e., non-informative prior).
Then, $m^{(1)}_{j \arrow f_{j,i}}(\bm \beta_j)$ being the product of Gaussian functions in (\ref{BPvs2f}) is also a Gaussian function \cite{Papoulis}.
Furthermore, based on the fact that the likelihood function $f_{j,i}$ is also Gaussian,
according to (\ref{BPf2vs}),  $m^{(1)}_{f_{j,i} \arrow  i}(\bm \beta_i)$ is a Gaussian
function.
Thus during each round of message exchange, all the messages are Gaussian functions and
only the mean vectors and covariance matrices need to be exchanged between neighboring factor nodes and variable nodes.

In general, for the $l^{th}$ ($l=2,3,\cdots$) round of message exchange, factor {{node} $f_{j,i}$ receives message $m^{(l)}_{ j \arrow f_{j,i}} (\bm \beta_j)$
in the form of $\mathcal{N}(\bm\beta_j|\bm v^{(l)}_{j\arrow f_{j,i}}, \bm C_{j\arrow f_{j,i}}^{(l)})$
from their neighboring variable nodes and then computes {{a} message using (\ref{BPf2vs}):
\begin{equation}\label{f2v}
\begin{split}
m^{(l)}_{f_{j,i} \arrow i}(\bm \beta_i)
&=  \int
m^{(l)}_{ j \arrow f_{j,i}}(\bm \beta_j)
f_{j,i}
d\bm \beta_j\\
&=
\int \mathcal{N}(\bm\beta_j|\bm v^{(l)}_{j\arrow f_{j,i}}, \bm C_{j\arrow f_{j,i}}^{(l)})
\mathcal{N}(\bm A_{i, j}\bm\beta_i|\bm A_{j, i}\bm\beta_j, \sigma_{i,j}^2\bm I_N)
 d\bm \beta_j.
\end{split}
\end{equation}
As the convolution of a pair of Gaussian function is also Gaussian function \cite{Papoulis},  after some algebraic manipulations, we obtain
$m^{(l)}_{f_{j,i} \arrow  i}(\bm \beta_i)
\propto \mathcal{N}(\bm\beta_i|\bm v^{(l)}_{f_{j,i}\arrow i}, \bm C_{f_{j,i}\arrow i}^{(l)})$, where the covariance matrix and mean vector are given by
\begin{equation}\label{messagecov}
\big[\bm C_{f_{j,i}\arrow i}^{(l)}\big]^{-1}
 =
\bm A_{i,j}^T
\bigg[ \sigma_{i,j}^2\bm I_N+ \bm A_{j,i}\bm C_{j\arrow f_{j,i}}^{(l)} \bm A_{j,i}^T \bigg]^{-1}
\bm A_{i,j},
\end{equation}
and
\begin{eqnarray}
\bm v^{(l)}_{f_{j,i}\arrow i}\label{f2vm}
=
\bm C_{f_{j,i}\arrow i}^{(l)}\bm A_{i, j}^T \bm A_{j, i}\bigg\{\bm A_{j, i}^T \bm A_{j, i}+
 \sigma_{i,j}^2\big[\bm C_{j\arrow f_{j,i}}^{(l)}\big]^{-1}\bigg\}^{-1}
\big[\bm C_{j\arrow f_{j,i}}^{(l)}\big]^{-1}
\bm v^{(l)}_{j\arrow f_{j,i}}.
\end{eqnarray}
On the other hand, using (\ref{BPvs2f}), the message passed from {{the} variable node to {{the} factor node is given by
the product of Gaussian distributions, which is
\begin{equation} \label{v2f}
\begin{split}
m^{(l)}_{j \arrow f_{j,i}}(\bm \beta_j)
&=  \!\!\!\!\prod_{f\in\B(\bm \beta_j)\setminus f_{j,i}} m^{(l-1)}_{ f \arrow j}(\bm \beta_j)\\
&\propto
\mathcal{N}(\bm\beta_j|\bm v^{(l)}_{j\arrow f_{j,i}}, \bm C^{(l)}_{j\arrow f_{j,i}}),
\end{split}
\end{equation}
where
\begin{eqnarray} \label{v2fV}
\big[\bm C^{(l)}_{j\arrow f_{j,i}}]^{-1}
=
\sum_{f\in\B(\bm \beta_j)\setminus f_{j,i}}
\big[\bm C_{f\arrow j}^{(l-1)}\big]^{-1}
\end{eqnarray}
and
\begin{eqnarray}\label{v2fm}
\bm v^{(l)}_{j\arrow f_{j,i}}&=& \bm C^{(l)}_{j\arrow f_{j,i}}
 \sum_{f\in\B(\bm \beta_j)\setminus f_{j,i}}
\big[\bm C_{f\arrow j}^{(l-1)}\big]^{-1}\bm v^{(l-1)}_{f\arrow j}.
\end{eqnarray}

Furthermore, during each round of message passing, each node can compute the belief for $\bm\beta_i$
as the product of all the incoming messages from neighboring factor nodes, which is given by
\begin{equation} \label{BPbelief}
b^{(l)}(\bm \beta_i)
 =  \prod_{ f\in\B(\bm \beta_i)} m^{(l-1)}_{ f \arrow i}(\bm \beta_i).
\end{equation}
According to (\ref{messagecov}), (\ref{f2vm}) and (\ref{BPbelief}), we can easily obtain
\begin{equation} \label{belief}
b^{(l)}(\bm\beta_i)\sim  \mathcal{N}\big(\bm\beta_i|
\big[\sum_{ f\in\B(\bm \beta_i)}\big[\bm C_{f\arrow i}^{(l-1)}\big]^{-1}\big]^{-1}
\sum_{ f\in\B(\bm \beta_i)}\big[\bm C_{f\arrow i}^{(l-1)}\big]^{-1}\bm v^{(l-1)}_{f\arrow i},
\big[\sum_{f\in\B(\bm \beta_i)}\big[\bm C_{f\arrow i}^{(l-1)}\big]^{-1}\big]^{-1}\big).
\end{equation}
Finally, the estimate of $\bm \beta_i$ in the $l^{th}$ iteration is
\begin{equation}\label{est1}
\hat {\bm \beta}_i^{(l)}
 =  \int \bm \beta_i b^{(l)}(\bm \beta_i)d\bm \beta_i
 = \big[\sum_{ f\in\B(\bm \beta_i)}\big[\bm C_{f\arrow i}^{(l-1)}\big]^{-1}\big]^{-1}
\sum_{ f\in\B(\bm \beta_i)}\big[\bm C_{f\arrow i}^{(l-1)}\big]^{-1}\bm v^{(l-1)}_{f\arrow i}.
\end{equation}

\subsection{Asynchronous Message Update}
In practical WSNs, there is neither factor nodes nor variable nodes.
These two kinds of {{messages} $m^{(l)}_{j \arrow f_{j,i}}(\bm \beta_j) $ and $m^{(l)}_{f_{j,i} \arrow i}(\bm \beta_i)$
are computed locally at node $j$, and only $m^{(l)}_{f_{j,i} \arrow i}(\bm \beta_i)$ is sent from node $j$ to node $i$ during each round of message exchange of BP.
Let $m^{(l)}_{j \arrow i}(\bm \beta_i)=\N(\bm \beta_i|\bm \gamma_{j\arrow i}^{(l)},\bm \Gamma_{j\arrow i}^{(l)})$ represent the physical message from node $j$ to node $i$.
Putting (\ref{v2fV}) and (\ref{v2fm}) into  (\ref{messagecov}) and (\ref{f2vm}), we have
\begin{equation} \label{f2fC}
\big[\bm \Gamma_{j\arrow i}^{(l)}\big]^{-1}
 =
\bm A_{i,j}^T
\bigg[  \sigma_{i,j}^2\bm I_N+ \bm A_{j,i}
\bigg[\bm \Gamma_j^{-1}
+\sum_{k\in\mathcal{I}(j)\setminus i} \big[\bm \Gamma_{k\arrow j}^{(l-1)}\big]^{-1}\bigg]^{-1}
 \bm A_{j,i}^T \bigg]^{-1}
\bm A_{i,j}
\end{equation}
and
\begin{eqnarray}\label{f2f2}
\bm \gamma^{(l)}_{j\arrow i}
&=&
\bm \Gamma_{j\arrow i}^{(l)}\bm A_{i,j}^T
\bigg[  \sigma_{i,j}^2\bm I_N+ \bm A_{j,i}
\bigg[\bm \Gamma_j^{-1}
+
\sum_{k\in\mathcal{I}(j)\setminus i} \big[\bm \Gamma_{k\arrow j}^{(l-1)}\big]^{-1}\bigg]^{-1}
 \bm A_{j,i}^T \bigg]^{-1} \\\nonumber
&& \times
\bm A_{j,i}
\bigg[\bm \Gamma_j^{-1}
+\sum_{k\in\mathcal{I}(j)\setminus i} \big[\bm \Gamma_{k\arrow j}^{(l-1)}\big]^{-1}\bigg]^{-1}
\bigg[ \bm \Gamma_j^{-1}\bm \gamma_j+
\sum_{k\in\mathcal{I}(j)\setminus i}
\big[\bm \Gamma_{k\arrow j}^{(l-1)}\big]^{-1}\bm \gamma^{(l-1)}_{k\arrow j}\bigg],
\end{eqnarray}
where $\bm \Gamma_j$ and $\bm \gamma_j$ are the covariance matrix and mean vector of
prior distribution of $\bm \beta_j$, respectively, and they
will never change during the updating process.

As shown in (\ref{f2fC}) and (\ref{f2f2}), from the perspective of node $j$,
the outgoing  message covariance $\bm \Gamma_{j\arrow i}^{(l)}$
and mean vector $\bm \gamma^{(l)}_{j\arrow i}$ computed by node $j$ at time $l$ depends on the incoming message covariance  $\bm \Gamma_{k\arrow j}^{(l-1)}$ and $\bm \gamma^{(l-1)}_{k\arrow j}$
from node $j$'s neighbour (i.e.,
$k\in \mathcal{I}(j)\setminus i$)  at time $l-1$.
However, in many situations, the inter-sensor
message exchange is possibly asynchronous due to random data packet
dropouts, and different nodes may update their messages at different frequencies.
If every node is allowed to update its belief only after receiving updated messages from all its neighbors, the convergence speed of the distributed algorithm would be slow.
Thus, some nodes should be allowed to update their beliefs more frequently than others,
as long as they receive some of the updates from their neighboring nodes
within a predetermined time period.
It means that when node $j$ computes $\bm \Gamma_{j\arrow i}^{(l)}$,
it may only have $\bm \Gamma_{k\arrow j}^{(s)}$ computed by node $k\in \mathcal{I}(j)\setminus i$ with
$s\leq l-1$.
In order to capture these asynchronous properties of message exchanges, we introduce the totally asynchronous model \cite{Bertsekas89} as follows.

Let the message covariance matrices and mean vectors available to node $j$ at time $l$ are $\bm \Gamma_{k\arrow j}^{(\tau^k_j(l-1))}$ and $\bm \gamma_{k\arrow j}^{(\tau^k_j(l-1))}$,
where $0 \leqslant\tau^k_j(l-1)\leqslant l-1$.
Without loss of generality, we assume that
node $j$ computes its outgoing messages to its neighboring nodes
according to a discrete time set $\mathcal{L}_j \subseteq \{0, 1, 2, \ldots \}$.
According to (\ref{f2fC}) and (\ref{f2f2}), the asynchronous message covariance and mean evolution {{are} defined as
\begin{eqnarray}\label{messagecov-outdate}
\big[\bm \Gamma_{j\arrow i}^{(l)}\big]^{-1}
 =
\left\{
\begin{array}{l}
\underbrace{\bm A_{i,j}^T
\bigg[  \sigma_{i,j}^2\bm I_N+ \bm A_{j,i}
\bigg[\bm \Gamma_j^{-1}
+\sum_{k\in\mathcal{I}(j)\setminus i} \big[\bm \Gamma_{k\arrow j}^{(\tau^k_j(l-1))}\big]^{-1}\bigg]^{-1}
 \bm A_{j,i}^T \bigg]^{-1}
\bm A_{i,j}}_{
\triangleq \mathbb{F}_{j\arrow i}\big(\bm \Gamma_j^{-1}+\sum_{k\in\mathcal{I}(j)\setminus i} \big[\bm \Gamma_{k\arrow j}^{(\tau^k_j(l-1))}\big]^{-1}\big)},
\quad \textrm{$l\in \mathcal{L}_j $},
\\
\big[\bm \Gamma_{j\arrow i}^{(l-1)}\big]^{-1}, \qquad \qquad \qquad \qquad \qquad \qquad \qquad \qquad \qquad \qquad \quad \textrm{otherwise},
\end{array}
\right.
\end{eqnarray}
and
\begin{eqnarray}\label{f2vm-outdate}
\bm \gamma^{(l)}_{j\arrow i}
=
\left\{
\begin{array}{l}
\bm \Gamma_{j\arrow i}^{(l)}\bm A_{i,j}^T
\bigg[ \sigma_{i,j}^2\bm I_N\!+\! \bm A_{j,i}
\bigg[\bm \Gamma_j^{-1}
\!+\!\sum_{k\in\mathcal{I}(j)\setminus i} \big[\bm \Gamma_{k\arrow j}^{(\tau^k_j(l-1))}\big]^{-1}\bigg]^{-1}
\bm A_{j,i}^T \bigg]^{-1}\bm A_{j,i}\\
\times
\!\bigg[\bm \Gamma_j^{-1}
\!+\!\sum_{k\in\mathcal{I}(j)\setminus i}\! \big[\bm \Gamma_{k\arrow j}^{(\tau^k_j(l-1))}\big]^{-1}\bigg]^{-1}
\!\bigg[\bm \Gamma_j^{-1}\bm \gamma_j\!+\!
\sum_{k\in\mathcal{I}(j)\setminus i}
\big[\bm \Gamma_{k\arrow j}^{(\tau^k_j(l-1))}\big]^{-1}\bm \gamma^{(\tau^k_j(l-1))}_{k\arrow j}\bigg],   \quad  \textrm{$l\in \mathcal{L}_j $}, \\
\bm \gamma^{(l-1)}_{j\arrow i}, \qquad \quad \qquad \qquad \qquad \qquad \qquad \qquad \qquad  \qquad \qquad \qquad\qquad \qquad\quad \textrm{otherwise}.
\end{array}
\right.
\end{eqnarray}
We assume $\lim_{l\rightarrow  \infty} \tau^k_j(l)= \infty$
for all $\{k, j\}\in \mathcal{E}$, which
guarantees that old information is eventually purged out of
the network, and that {{each node eventually exchanges messages with its neighboring nodes}.

The asynchronous iterative algorithm is summarized as follows.
The algorithm is started by setting the messages from node $j$ to node $i$ as $m^{(0)}_{j \arrow  i}(\bm \beta_i)=\N(\bm\beta_i;\bm 0, +\infty\bm I_2)$
\footnote{{Since the message updating using (\ref{messagecov-outdate}) and (\ref{f2vm-outdate}) only involves inverse
of covariance matrix, in practice, we can set the inverse of the initial covariance matrix as 0.}}.
Each node $i$  computes its outgoing message according to (\ref{messagecov-outdate}) and (\ref{f2vm-outdate}) at independent time $l\in\mathcal{L}_i$ with its available
$\big[\bm \Gamma_{j\arrow i}^{(\tau^j_i(l-1))}\big]^{-1}$ and $\bm \gamma_{j\arrow i}^{(\tau^j_i(l-1))}$.
The corresponding belief of node $i$ at  time $l$ is computed as
\begin{equation} \label{belief2}
b^{(l)}(\bm\beta_i)\sim  \mathcal{N}\big(\bm\beta_i|
\bm \mu_i^{(l)}, \bm P_i^{(l)}\big),
\end{equation}
where the belief covariance matrix is
\begin{equation} \label{beliefcov}
\bm P_i^{(l)} =
\bigg[\bm \Gamma_i^{-1}
+\sum_{j\in\mathcal{I}(i)} \big[\bm \Gamma_{j\arrow i}^{(\tau^j_i(l-1))}\big]^{-1}\bigg]^{-1},
\end{equation}
and mean vector is
\begin{equation} \label{beliefmean}
\bm \mu_i^{(l)}=\bm P_i^{(l)}\bigg[
\bm \Gamma_i^{-1}\bm \gamma_i+
\sum_{j\in\mathcal{I}(i)}
\big[\bm \Gamma_{j\arrow i}^{(\tau^j_i(l-1))}\big]^{-1}\bm \gamma^{(\tau^j_i(l-1))}_{j\arrow i}\bigg].
\end{equation}

The iterative computation terminates when (\ref{belief2}) converges or the maximum number of {{iterations} is reached.
Then each sensor computes
its clock skew and offset according to
\begin{equation}\label{skewoffset}
\hat{\alpha}_i  = {1}/{\bm \mu_i^{(l)}(1) },  \quad
\hat{\theta}_i  =  {\bm \mu_i^{(l)}(2) }/{\bm \mu_i^{(l)}(1) },
\end{equation}
where $\bm \mu_i^{(l)}(k)$ denotes the $k^{th}$ element of $\bm \mu_i^{(l)}$.
\section{Asynchronous BP Convergence Analysis}\label{Section 4}
It is important to note that the BP message {{updates (\ref{f2v}) and (\ref{v2f}) are specially designed for the computation} of marginal functions (e.g., $g_i(\bm \beta_i)$ in (\ref{marginal})) on cycle-free FG and
it is known that the beliefs will converge to the exact marginal functions.
On the other hand, the
BP algorithm may be applied to FG
with cycles, but since messages will be passed multiple times on a
given edge, no convergence can be guaranteed \cite{DiagnalDominant}.
Although some of the most exciting applications of BP
algorithm like the decoding of turbo codes and low-density
parity-check codes \cite{Kschischang} do not exhibit divergence in the simulations even under loopy FG,
{{there are still many applications where BP do diverge.}
General  sufficient condition for convergence of loopy FGs is available in \cite{WalkSum1}
but it requires the knowledge of the joint posterior
distribution of all unknown variables as shown in the integrand of (\ref{marginal}), and
is difficult to verify for large-scale dynamic networks.
{{Reference \cite{LengMei11BP} proved the convergence of BP in the context of distributed  clock offset synchronization, by exploiting the Perron-Frobenius theorem in the context of matrices with nonnegative elements.}
However, in {{the} vector variable case (both clock skew and offset),
the BP message covariance matrices contain negative elements, and the analysis in \cite{LengMei11BP} is not applicable.  Besides, the effect of asynchronous message-update was not addressed in \cite{LengMei11BP}.
In the following, we will prove the convergence of asynchronous vector BP messages in distributed clock synchronization.

Defining the operator $\mathbb{F}_{j\arrow i}(\cdot)$  corresponding to the update of the message covariance in (\ref{messagecov-outdate}),
the following properties are first established.

{\textbf{Lemma 1.}} The updating operator $\mathbb{F}_{j\arrow i}(\cdot)$ satisfies the following properties:

\noindent Property  \romannumeral 1): $\mathbb{F}_{j\arrow i}(\bm 0) = \bm 0$.

\noindent Property  \romannumeral 2): $\mathbb{F}_{j\arrow i}(\bm X) \succ 0$, if $\bm X \succ \bm 0$.

\noindent Property  \romannumeral 3):
$\mathbb{F}_{j\arrow i}(\bm X) \succeq \mathbb{F}_{j\arrow i}(\bm Y)$, if $\bm X  \succeq \bm Y\succ \bm 0$.

\noindent  \textit{Proof}:
\noindent  Property  \romannumeral 1) is apparent according to (\ref{messagecov-outdate}).
The proof of property \romannumeral 2) is given as follows.
Let $\bm X \succ 0$, it is obvious that
 $\bm X^{-1}\succ \bm 0$,
which means $\bm y^T\bm X^{-1}\bm y \geq 0$ for any $\bm y$.
Putting $\bm y=\bm A_{j,i}^T\bm x$, we have
$\bm x^T\bm A_{j,i}\bm X^{-1}\bm A_{j,i}^T\bm x\geq \bm 0$.
As sum of positive definite and positive semi-definite matrices is positive definite, we have
$\big[  \sigma_{i,j}^2\bm I_N+ \bm A_{j,i}\bm X^{-1}  \bm A_{j,i}^T \big]^{-1}
\succ \bm 0$.
Since $\bm A_{i,j}$  is of full column rank, we obtain
$\bm A_{i,j}^T\big[  \sigma_{i,j}^2\bm I_N+ \bm A_{j,i}\bm X^{-1}  \bm A_{j,i}^T \big]^{-1}\bm A_{i,j}
\succ \bm 0$.
Thus, property \romannumeral 2) is proved.
For the proof of property  \romannumeral 3),
let $\bm X  \succeq \bm Y\succ \bm 0$, then
we have $\bm Y^{-1}  - \bm X^{-1}\succ \bm 0$ \cite{MatrixTricks},
which means $\bm y^T(\bm Y^{-1}  - \bm X^{-1})\bm y \geq 0$ for any $\bm y$.
Let $\bm y=\bm A_{j,i}^T\bm x$, we have
$\bm x^T\bm A_{j,i}\bm Y^{-1}\bm A_{j,i}^T\bm x\geq \bm x^T\bm A_{j,i}\bm X^{-1}\bm A_{j,i}^T\bm x$.
Hence, we have
$\big[  \sigma_{i,j}^2\bm I_N+ \bm A_{j,i}\bm X^{-1}  \bm A_{j,i}^T \big]^{-1}
\succeq  \big[  \sigma_{i,j}^2\bm I_N+ \bm A_{j,i}\bm Y^{-1}  \bm A_{j,i}^T \big]^{-1}$.
Due to the fact that $\bm A_{i,j}$  is of full column rank, we have
$\bm A_{i,j}^T\big[  \sigma_{i,j}^2\bm I_N+ \bm A_{j,i}\bm X^{-1}  \bm A_{j,i}^T \big]^{-1}\bm A_{i,j}
\succeq \bm A_{i,j}^T\big[  \sigma_{i,j}^2\bm I_N+ \bm A_{j,i}\bm Y^{-1}  \bm A_{j,i}^T \big]^{-1}\bm A_{i,j}$,
which is equivalent to  $\mathbb{F}_{j\arrow i}(\bm X) \succeq \mathbb{F}_{j\arrow i}(\bm Y)$.
\QEDA

To consider the updates of all message covariance matrices, we {{introduce} the following definitions.
Let
${\bf\Xi}^{(\tau(l-1))}\triangleq \big[[\bm \Gamma_{1\arrow k}^{(\tau^1_k(l-1))}]^{-1};\ldots;[\bm \Gamma_{j\arrow i}^{(\tau^j_i(l-1))}]^{-1};\ldots;[\bm \Gamma_{r\arrow M}^{(\tau^r_M(l-1))}]^{-1};
\bm \Gamma_{1}^{-1};\ldots;\bm \Gamma_{M}^{-1}\big]$
be the collection of all available message covariance (including prior covariance) matrices in the network at time $l$,
and
${\bf\Xi}^{(l)}\triangleq \big[[\bm \Gamma_{1\arrow k}^{(l)}]^{-1};\ldots;
[\bm \Gamma_{j\arrow i}^{(l)}]^{-1};\ldots;[\bm \Gamma_{r\arrow M}^{(l)}]^{-1}\big]$ be the collection of all outgoing message covariances in the network at time $l$.
Define ${\bf\Xi}^{(l)}\succeq^b \bm 0 $ if its component $[\bm \Gamma_{j\arrow i}^{(l)}]^{-1}\succeq \bm 0$; and
${\bf\Xi}^{(l)}\succeq^b {\bf\Xi}^{(l-1)}$ if their corresponding components satisfy
$[\bm \Gamma_{j\arrow i}^{(l)}]^{-1}\succeq [\bm \Gamma_{j\arrow i}^{(l-1)}]^{-1}$.
The same  definitions apply to ${\bf\Xi}^{(\tau(l))}$.
Furthermore, we define the function $\mathbb{F}\triangleq(\mathbb{F}_{1\arrow k}, \ldots, \mathbb{F}_{j\arrow i}, \ldots, \mathbb{F}_{r\arrow M})$ which satisfies
${\bf\Xi}^{(l+1)} = \mathbb{F}({\bf\Xi}^{(\tau(l))}) $.
Then we have the following lemma.

{\textbf{Lemma 2.}}
${\bf\Xi}^{(l)}$ and ${\bf\Xi}^{(\tau(l-1))}$ satisfy the following properties:

\noindent Property  \romannumeral 4):
If ${\bf\Xi}^{(l)}\succeq^b {\bf\Xi}^{(l-1)}$, then ${\bf\Xi}^{(\tau(l))}\succeq^b {\bf\Xi}^{(\tau(l-1))}$.

\noindent Property  \romannumeral 5):
If ${\bf\Xi}^{(\tau(l))}\succeq^b {\bf\Xi}^{(\tau(l-1))}$, then
$\mathbb{F}({\bf\Xi}^{(\tau(l))})\succeq^b \mathbb{F}({\bf\Xi}^{(\tau(l-1))})$ or equivalently
${\bf\Xi}^{(l+1)}\succeq^b {\bf\Xi}^{(l)}$.

\noindent
\noindent  \textit{Proof}:
The proofs of properties  \romannumeral 4) and  \romannumeral 5) rest on the basic definitions
that $[\bm \Gamma_{j\arrow i}^{(l)}]^{-1}$ represents the message covariance matrix sends from
node $j$ to node $i$ at time $l$, and
$[\bm \Gamma_{j\arrow i}^{(\tau(l))}]^{-1}$ represents message covariance matrix received by node $i$ at time
$l$.
If $[\bm \Gamma_{j\arrow i}^{(l)}]^{-1} \succeq [\bm \Gamma_{j\arrow i}^{(l-1)}]^{-1}$, it is obvious that the received covariance will satisfy
$[\bm \Gamma_{j\arrow i}^{(\tau(l))}]^{-1} \succeq [\bm \Gamma_{j\arrow i}^{(\tau(l-1))}]^{-1}$.
Since ${\bf\Xi}^{(l)}$ and
${\bf\Xi}^{(\tau(l))}$  contain $[\bm \Gamma_{j\arrow i}^{(l)}]^{-1}$ and
$[\bm \Gamma_{j\arrow i}^{(\tau(l))}]^{-1}$  as components respectively,
property  \romannumeral 4) is obvious.
On the other hand, property \romannumeral 5) is apparent since each of the corresponding components in
${\bf\Xi}^{(\tau(l))}$ and ${\bf\Xi}^{(\tau(l-1))}$  satisfies property \romannumeral 1) or  \romannumeral 3)  in Lemma 1.
\QEDA

Now we present the convergence property of the covariance matrix in
the local beliefs.

{\textbf{Theorem 1.}}
{{For the totally asynchronous clock synchronization algorithm, the covariance matrix $\bm P_i^{(l)}$ of belief $b_i^{(l)}(\bm \beta_i)$ at each node converges to a positive definite matrix regardless of network topology.}

\noindent  \textit{Proof}:
Initially, all messages are non-informative, that is, $\bm \Gamma_{j\arrow i}^{\tau(-1)}=\bm \Gamma_{j\arrow i}^{(0)}=
\infty\bm I_2$.
From (\ref{messagecov-outdate}), properties  \romannumeral 1) and \romannumeral 2), we obtain that
$\big[\bm \Gamma_{j\arrow i}^{(l)}\big]^{-1}\succ \bm 0$ only if $\bm \Gamma_j^{-1}
+\sum_{k\in\mathcal{I}(j)\setminus i} \big[\bm \Gamma_{k\arrow j}^{(\tau^k_j(l-1))}\big]^{-1} \succ \bm 0$.
Therefore, the first batch of nodes having outgoing covariance $\big[\bm \Gamma_{j\arrow i}^{(l)}\big]^{-1}\succ \bm 0$
must have $\bm \Gamma_{j}^{-1}\succ \bm 0$, i.e., informative prior.
Let the first message updating event in the network occurs at time $s$.
We have
${\bf\Xi}^{(s)}\succeq^b {\bf\Xi}^{(s-1)}$.
Applying  property \romannumeral 4),  we further obtain ${\bf\Xi}^{\tau(s)}\succeq^b {\bf\Xi}^{\tau(s-1)}$.

Suppose
${\bf\Xi}^{(\tau(l))}\succeq^b{\bf\Xi}^{(\tau(l-1))}$ for $l\geq s$,
according to property \romannumeral 5),
${\bf\Xi}^{(l+1)}\succeq^b {\bf\Xi}^{(l)}$.
Thus $ {\bf\Xi}^{(\tau(l+1))}\succeq^b {\bf\Xi}^{(\tau(l))}$ for $l\geq s$ due to property \romannumeral 4).
Hence, by induction the updating relationship of ${\bf\Xi}^{(\tau(l))}$ is
\begin{equation}\label{Cseq2}
 \ldots \succeq^b {\bf\Xi}^{(\tau(l))}
 \ldots \succeq^b {\bf\Xi}^{(\tau(s))}
\succeq^b \bm 0.
\end{equation}
Focusing on node $i$,  we obtain
\begin{equation} \label{C1}
\ldots \succeq
\bm \Gamma_i^{-1}+\sum_{j\in\I(i)}\big[\bm \Gamma_{j\arrow i}^{(\tau^j_i(l))}\big]^{-1}
\ldots \succeq
\bm \Gamma_i^{-1}+\sum_{j\in\I(i)}\big[\bm \Gamma_{j\arrow i}^{(\tau^j_i(s))}\big]^{-1}.
\end{equation}
Since {{a} strongly connected network is considered, there must be one of $[\bm \Gamma_{j\arrow i}^{(\tau^j_i(l^{\prime}-1))}]^{-1}\succ \bm 0$  for some $l^{\prime}\geq s$, and therefore (\ref{C1}) is lower bounded by the all-zero matrix.
{{Furthermore, since $\infty\bm I_2\succeq \bm \Gamma_j^{-1}+\sum_{k\in\mathcal{I}(j)\setminus i} \big[\bm \Gamma_{k\arrow j}^{(\tau^k_j(l-1))}\big]^{-1}$,
according to property \romannumeral 3),
$\mathbb{F}_{j\arrow i}\big(\infty\bm I_2\big)\succeq
\mathbb{F}_{j\arrow i}\big(\bm \Gamma_j^{-1}+\sum_{k\in\mathcal{I}(j)\setminus i} \big[\bm \Gamma_{k\arrow j}^{(\tau^k_j(l-1))}\big]^{-1}\big)$.
Using the definition of $\mathbb{F}_{j\arrow i}(\cdot)$ in (\ref{messagecov-outdate}), this is equivalent to
$\frac{1}{\sigma_{i,j}^2}\bm A_{i, j}^T\bm A_{i, j}\succeq
\big[\bm \Gamma_{j\arrow i}^{(\tau^j_i(l))}\big]^{-1}$.
Therefore, we can add an upper bound to (\ref{C1}) and obtain}
\begin{equation} \label{C2}
\bm \Gamma_i^{-1}+\sum_{j\in\I(i)}\frac{1}{\sigma_{i,j}^2}\bm A_{i, j}^T\bm A_{i, j}
\succeq \ldots \succeq
\bm \Gamma_i^{-1}+\sum_{j\in\I(i)}\big[\bm \Gamma_{j\arrow i}^{(\tau^j_i(l^{\prime}+1))}\big]^{-1}
\succeq
\bm \Gamma_i^{-1}+\sum_{j\in\I(i)}\big[\bm \Gamma_{j\arrow i}^{(\tau^j_i(l^{\prime}))}\big]^{-1}
\succ \bm 0.
\end{equation}
Then, applying matrix inverse to (\ref{C2}) and using the definition of $ \bm P_i^{(l)}$ in (\ref{beliefcov}) {{results} in
\begin{equation} \label{P1}
 \bm P_i^{(l^{\prime})}  \succeq    \bm P_i^{(l^{\prime}+1)}\succeq \ldots \succeq \big[\bm \Gamma_i^{-1}+\sum_{j\in\I(i)}\frac{1}{\sigma_{i,j}^2}\bm A_{i, j}^T\bm A_{i, j}\big]^{-1} \succ \bm 0,
\end{equation}
where the inequality relationship is due to the fact that if $\bm X, \bm Y \succ 0$  and $\bm X  \succeq \bm Y$, then $\bm Y^{-1}\succeq \bm X^{-1}$ \cite{MatrixTricks}.
Consequently, such  non-increasing positive definite matrix sequence $ \bm P_i^{(l)}$ in (\ref{P1}) converges to a positive definite matrix \cite{MatrixConverge}.
\QEDA

The importance of Theorem $1$ is that the covariance matrices of belief
always converge regardless of network topology as long as informative prior exists.
Next, we show the convergence of belief mean vectors.

{\textbf{Theorem 2.}}
For the totally asynchronous belief propagation, the mean vector $\bm\mu_i^{(l)}$ of the belief $b^{(l)}(\bm \beta_i)$ converges to a constant vector regardless of the network topology.

\noindent  \textit{Proof}:
From (\ref{Cseq2}) in the proof of Theorem 1, we can readily see that
$\bm \Gamma_{k\arrow j}^{(\tau^k_j(l))}$  satisfies:
$\ldots \succeq  [\bm \Gamma_{k\arrow j}^{(\tau^k_j(l))}]^{-1}\succeq \ldots \succeq [\bm \Gamma_{k\arrow j}^{(\tau^k_j(s))}]^{-1}\succeq \bm 0$.
If there is a path from any {{node} with informative prior to node $k$,
according to property \romannumeral 2), there must be a time instant $l^{\prime}$ after which
$\ldots \succeq  [\bm \Gamma_{k\arrow j}^{(\tau^k_j(l^{\prime}+1))}]^{-1}\succeq \ldots \succeq [\bm \Gamma_{k\arrow j}^{(\tau^k_j(l^{\prime}))}]^{-1}\succ\bm 0$.
Hence $\bm \Gamma_{k\arrow j}^{(\tau^k_j(l^{\prime}))}$ is convergent \cite{MatrixConverge}.
On the other hand,
if there is no path from any {{node} with informative prior to node $k$, we have
$\ldots=[\bm \Gamma_{k\arrow j}^{(\tau^k_j(l))}]^{-1}=\ldots =[\bm \Gamma_{k\arrow j}^{(\tau^k_j(0))}]^{-1}=\bm 0$.
Either case implies $\bm \Gamma_{k\arrow j}^{(\tau^k_j(l))}$ converges to a  matrix
$\bm \Gamma_{k\arrow j}^{(\ast)}$.
From (\ref{messagecov-outdate}), if $\bm \Gamma_{k\arrow j}^{(\tau^k_j(l))}$ converges, we have
$\bm \Gamma_{j\arrow i}^{(l)}$ also {{converges to a fixed matrix}
$\bm \Gamma_{j\arrow i}^{(\ast)}$.
Then, (\ref{f2vm-outdate}) can be rewritten as
\begin{eqnarray}\label{f2f2-2}
\bm \gamma^{(l)}_{j\arrow i}
=
\left\{
\begin{array}{l}
\bm \Gamma_{j\arrow i}^{(\ast)}\bm A_{i,j}^T\bigg[  \sigma_{i,j}^2\bm I_N+ \bm A_{j,i}
\bigg[\bm \Gamma_{j}^{-1}
+\sum_{k\in\I(j)\setminus i} \big[\bm \Gamma_{k\arrow j}^{(\ast)}\big]^{-1}\bigg]^{-1}
 \bm A_{j,i}^T \bigg]^{-1} \bm A_{j,i}\\
\times
\bigg[\bm \Gamma_{j}^{-1}
+\sum_{k\in\I(j)\setminus i} \big[\bm \Gamma_{k\arrow j}^{(\ast)}\big]^{-1}\bigg]^{-1}
\bigg[\bm \Gamma_j^{-1}\bm \gamma_j+
\sum_{k\in\mathcal{I}(j)\setminus i}
\big[\bm \Gamma_{k\arrow j}^{(\ast)}\big]^{-1}\bm \gamma^{(\tau^k_j(l-1))}_{k\arrow j}\bigg],  \quad  \textrm{$l\in \mathcal{L}_j $}, \\
\bm \gamma^{(l-1)}_{j\arrow i}, \qquad \qquad \qquad \qquad \qquad \qquad \qquad \qquad  \qquad \qquad \qquad \qquad\qquad \quad \textrm{otherwise}.
\end{array}
\right.
\end{eqnarray}
Without loss of generality, define $\bm \gamma^{(l)}$ as a vector containing all $\bm \gamma_{j}$ and outgoing message mean $\bm \gamma^{(l)}_{j\arrow i}$
with ascending index first on $j$ and then on $i$ ($\bm \gamma_{j}$ can be interpreted as $\bm \gamma_{j\rightarrow j}$ for the ordering),
and $\bm \gamma^{(l-1)}$ is the vector constituted by  available message means with the same ordering.
It should be noticed that the order of $\bm \gamma^{(l)}_{j\arrow i}$ arranged in $\bm \gamma^{(l)}$ can be arbitrary as long as it does not change after the order is fixed.
Then, (\ref{f2f2-2}) can be expressed as
\begin{equation} \label{vectoriter-x2-1}
\bm \gamma^{(l)} = \bm Q^{(l)} \bm \gamma^{(l-1)},
\end{equation}
where the specific structure of $ \bm Q^{(l)}$ depends on the messages sent and received at time $l$.
Notice that $ \bm Q^{(l)}$ is time-varying due to asynchronous updating.
The convergence condition for the asynchronous system (\ref{vectoriter-x2-1}) turns out to be related to the system matrix of the corresponding synchronous system [32, p. 434], [33, p. 14].
Consider $\mathcal{L}_j=\{0,1,2, \ldots\}$ for all $j=1,2,\ldots, M$,
the asynchronous system (\ref{vectoriter-x2-1}) becomes a synchronous one:
\begin{equation} \label{vectoriter-x2}
\bm \gamma^{(l)} = \bm Q \bm \gamma^{(l-1)},
\end{equation}
where $\bm Q$ is now independent of iteration number $l$.
The necessary and sufficient convergence condition for the asynchronous iteration (\ref{vectoriter-x2-1}) is
$ \rho(|\bm Q|)<1$ [32, p. 434], where $|\bm Q|$  denotes the matrix {{whose  elements are the absolute values of those in $\bm Q$.}
Next, we prove that $ \rho(|\bm Q|)<1$.

First, construct the new linear iteration as
\begin{equation}\label{vectoriter-x3}
{\bm x}^{(r)} = \tilde{\bm Q} {\bm x}^{(r-1)},
\end{equation}
where $\tilde{\bm Q}=|\bm Q|$, ${\bm x}^{(r)} $ is a vector with the same structure as
$\bm \gamma^{(r)}$ and ${\bm x}^{(0)}=\bm \gamma^{(0)}$.
Since there is always a positive value $\eta$, satisfying $\eta>\sum_{i\neq j}|[\tilde{\bm Q}]_{i,j}|$ for all $i$,
we have $\eta\bm I+\tilde{\bm Q}$ is strictly diagonally dominant and then $\eta\bm I+\tilde{\bm Q}$ is nonsingular \cite{MatrixHorn}.
Hence, {{the} arbitrary initial value ${\bm x}^{(0)}$ can be expressed in terms of the eigenvectors of $\eta\bm I+ \tilde{\bm Q}$
as ${\bm x}^{(0)}= \sum_{d=1}^{D}c_{d}\bm q_{d}$,
where $D$ is the dimension of matrix $\tilde{\bm Q}$ and $\bm q_1$, $\bm q_2$,$\cdots$, $\bm q_D$ are the eigenvectors of $\eta\bm I+\tilde{\bm Q}$.
Since the eigenvectors of $\eta\bm I+\tilde{\bm Q}$ {{are} the same as {{those} of $\tilde{\bm Q}$, and the eigenvalues of $\eta\bm I+\tilde{\bm Q}$ are $\eta+\lambda_d$ ($1\leqslant d \leqslant D$), where $\lambda_d$ is the eigenvalue of $\tilde{\bm Q}$, we have
\begin{eqnarray} \label{eigen}
{{\bm x}}^{(r)}=
\tilde{\bm Q}^r{{\bm x}}^{(0)}
= \sum_{{d}=1}^{D} {c_{d}}{\lambda^{r}_{d}}\bm q_d.
\end{eqnarray}
Without loss of generality, suppose ${\lambda}_{d}$ are arranged in  descending order as
\begin{equation}
|{\lambda}_1|\geq|{\lambda}_2|\geq  \cdots \geq|{\lambda}_{D}|.
\end{equation}
Let the eigenvalue with the largest magnitude has a multiplicity of  $d_0$.
Then ${{\lambda}_{{d}}}/{{\lambda}_{1}}<1$ for ${d}>{d}_0$
and $({{\lambda}_{{d}}}/{{\lambda}_{1}})^r=0$ if $r$ is large enough.
We then obtain
\begin{eqnarray} \label{eigen2}
{\lim_{r\rightarrow \infty}}{{\bm x}}^{(r)}
= {\lambda}_1^{r} \sum_{{d}=1}^{{d}_0}c_{d}\bm q_{d}.
\end{eqnarray}

On the other hand, putting $j=1$ into (\ref{messagecov-outdate}), and  noting $\bm \Gamma_1^{-1}=\infty\bm I_2$, we obtain
$[\bm \Gamma_{1\arrow i}^{(l)}]^{-1}=\frac{1}{\sigma_{i,1}^2}\bm A_{i, 1}^T\bm A_{i, 1}$, for $l\in \mathcal{L}_i$.
But since this outgoing covariance from {{the} reference node is independent of time $l$, we can combine the two cases in (\ref{messagecov-outdate}).
Substituting this result into (\ref{f2vm-outdate}), we have
${\bm \gamma}^{(l)}_{1\arrow i}
=
\frac{1}{\sigma^2_{i,1}} \big[\bm A_{i,1}^T \bm A_{i,1}\big]^{-1}\bm A_{i,1}^T
\bm A_{1,i}\bm \beta_1$ ($i\neq 1$),
which shows that ${\bm \gamma}^{(l)}_{1\arrow i}$ is also independent of time  $l$.
Consequently, according to (\ref{vectoriter-x2}),  ${\bm \gamma}^{(l)}_{1\arrow i}=[\bm Q]_{1:2,1:D}{\bm \gamma}^{(l-1)}$
and $[\bm Q]_{1:2,1:D}=[\bm I_2, \bm 0]$.
Hence, $|[\bm Q]_{1:2,1:D}|\bm x^{(0)}={\bm x}^{(0)}_{1\arrow i}
={\bm x}^{(1)}_{1\arrow i}$.
In general, we also have ${\bm x}^{(r)}_{1\arrow i}={\bm x}^{(0)}_{1\arrow i}$ for all $r$.
Therefore, we can put ${{{\bm x}}}^{(r)}(m_i)={\bm \gamma}^{(l)}_{1\arrow i}\triangleq \xi_c$ being a constant into
(\ref{eigen2}) to obtain
${\lambda}_1^{r} = \frac{\xi_c}{\sum_{{d}=1}^{{d}_0} c_{d}\bm q_{d}(m_i) }$
for $r$ large enough.
Substituting it back into (\ref{eigen2}) yields
\begin{equation}
{\lim_{r\rightarrow \infty}} {{\bm x}}^{(r)} = \frac{\xi_c\sum_{{d}=1}^{{d}_0}c_{d}\bm q_{d}}{\sum_{{d}=1}^{{d}_0} c_{d}\bm q_{d}(m_i) }.
\end{equation}
It is obvious that ${{\bm x}}^{(r)} $ does not change when $r$ is large enough, and
therefore, ${{\bm x}}^{(r)}$ in (\ref{vectoriter-x3}) converges.
Hence, the spectrum radius $\rho(\tilde{\bm Q})=\rho({|\bm Q|})<1$ \cite{Demmel}, and
according to  [32, p. 434], the asynchronous version of the iteration  given by (\ref{vectoriter-x2-1})
converges.
Finally, with $\bm \mu_i^{(l)}$ defined in (\ref{beliefmean}), since
$ \bm P_{i}^{(l)}$, $\bm \Gamma_{j\arrow i}^{(l)}$ and $\bm \gamma^{(l)}_{j\arrow i}$
converge, we can draw the conclusion that the
vector sequence $\{\bm\mu_i^{(1)}, \bm\mu_i^{(2)}, \ldots\}$
converges.
\QEDA

Theorems 1 and 2 reveal that the BP messages converge.  Next, we address how good is the clock parameters estimate
(\ref{skewoffset}) based on the converged message mean $ \bm \mu_i^{\ast} = \lim_{l \to +\infty} \bm \mu_i^{(l)}$.
Since the prior $p(\bm \beta_i)$ and likelihood function $p(\bm A_{i,j}, \bm A_{j,i}|\bm \beta_i,\bm \beta_j)$
are both Gaussian distribution and it is known that if  Gaussian BP (synchronous or asynchronous) converges, the means of the beliefs computed by BP equal the means of the marginal posterior distribution \cite{WalkSum1,ConsensusPro}, i.e.,
$\bm \mu_i^{\ast}=\hat{\bm \beta}^{\text{MMSE}}_i\triangleq \int \cdots \int \bm \beta_i
p\big(\bm \beta_1, \bm \beta_2, \ldots, \bm \beta_M |{\{{\bm A}_{i, j}\}}_{\{i,j\}\in \mathcal{E}}\big) d\bm \beta_2\cdots d\bm \beta_M$.
Stacking $\hat{\bm \beta}^{\text{MMSE}}_i$ into a block vector $\hat{\bm \beta}^{\text{MMSE}}=[(\hat{\bm \beta}^{\text{MMSE}}_2)^T,\ldots,(\hat{\bm \beta}^{\text{MMSE}}_M)^T]^T$ gives
\begin{equation} \label{centralizedMMSE}
\begin{split}
\hat{\bm \beta}^{\text{MMSE}}
=   \int...\int
{[\bm \beta^T_2, \ldots, \bm \beta^T_M]^T}
p\big(\bm \beta_1,\bm \beta_2,\ldots,\bm \beta_M |{\{{\bm A}_{i, j}\}}_{\{i,j\}\in \mathcal{E}}\big)  d\bm \beta_2\ldots d\bm \beta_{M}.
\end{split}
\end{equation}
It is obvious that $\bm \mu^{\ast}=\big[(\bm\mu_2^*)^T, \ldots , (\bm\mu_M^*)^T\big]^T$ equals the centralized joint MMSE estimator $\hat{\bm \beta}^{\text{MMSE}}$.
In case of non-informative prior, $ \hat{\bm\beta}^{\textrm{MMSE}}$ is the mean of the joint likelihood function.
Since the mean and maximum of a Gaussian distribution are the same,
$\bm \mu^{\ast}$ equals the centralized joint maximum likelihood (ML) estimator under non-informative prior.

{\textbf{Theorem 3.}}
Under non-informative prior of $\bm\beta_i$, the MSE of the estimator
$[ \frac{1}{\bm\mu_2^{\ast}(1)},\frac{\bm\mu_2^{\ast}(2)}{\bm\mu_2^{\ast}(1)},\ldots, \frac{1}{\bm\mu_M^{\ast}(1)}\\,\frac{\bm\mu_M^{\ast}(2)}{\bm\mu_M^{\ast}(1)}]^T$ obtained from the converged BP message mean vectors
$\bm \mu_i^{\ast}$ asymptotically approaches the centralized CRB of $\bm \zeta=[\theta_{2}, \alpha_{2}, \ldots, \theta_{M}, \alpha_{M} ]^T$, where the CRB is given by (\ref{CRBreal}) in the Appendix.

\noindent  \textit{Proof}:
As discussed after (\ref{centralizedMMSE}), under non-informative prior, $\bm\mu^{\ast}$ equals the centralized joint ML estimator
of $[{\bm \beta}^T_2,\ldots, \bm \beta^T_M]^T$.
Due to $\bm\beta_i=[\frac{1}{\alpha_i} , \frac{\theta_i}{\alpha_i}]^T$ and from
the invariance property of ML estimator \cite{Kay},
$[ \frac{1}{\bm\mu_2^{\ast}(1)},\frac{\bm\mu_2^{\ast}(2)}{\bm\mu_2^{\ast}(1)},\ldots, \frac{1}{\bm\mu_M^{\ast}(1)},\frac{\bm\mu_M^{\ast}(2)}{\bm\mu_M^{\ast}(1)}]^T$
is the ML estimator of $\bm \zeta=[\theta_{2}, \alpha_{2}, \ldots, \theta_{M}, \alpha_{M} ]^T$,
with the corresponding MSE asymptotically approaches
 the centralized CRB of $\bm \zeta$ derived in (\ref{CRBreal}) in the Appendix.
\QEDA

Synchronous message updating, i.e., $\mathcal{L}_{1}=\ldots = \mathcal{L}_{M}$
and $\tau^{k}_{j}(l-1)=l-1$, is obviously a special case of   (\ref{messagecov-outdate}) and (\ref{f2vm-outdate}).
Hence, Theorem 1,  Theorem 2 and Theorem 3 also apply to the synchronous BP.

\section{Simulation Results}\label{Section 5}
This section presents numerical results to assess the performance of the proposed algorithm.
Simulation results of estimation mean-square-error (MSE) are presented for random networks with
$25$ nodes randomly located in an area of  size $[0, 300]\times[0, 300]$.
Each node can only communicate with the sensor nodes that are within its radio range,
which is assumed to be 90.
In each simulation, clock skews $\alpha_i$ and clock offsets $\theta_i$ are uniformly distributed in the range $[-0.945, 1.055]$ and $[-5.5,5.5]$, respectively.
The fixed delay $d_{i,j}$ is uniformly distributed in $[8, 12]$
and variance of random delay $\sigma^2_i=0.05$ is assumed to be identical for all nodes.
{{$5000$ Monte-carlo simulation trials were performed to obtain the average performance of each point in all the figures presented in this section.}
Without loss of generality, Node $1$ is selected as the reference node with $\bm\beta_1=[1,0]^T$, and
$p(\bm\beta_1)= \delta(\bm \beta_1-[1, 0]^T)$.
For {{the} other nodes, non-informative prior is assumed $p(\bm\beta_i)= \N(\bm\beta_i;\bm 0, +\infty\bm I_2)$.
The probability of node $i$ successfully pass a message to its direct neighboring node $j$ is
$p_{i,j}$ for $\{i, j\}\in \mathcal{E}$.
With $p_{i, j}\neq 1$, we can emulate an asynchronous network.
To serve as a reference of the distributed estimation performance,
the CRB for centralized estimation is derived
in the Appendix.

Fig. \ref{25NodeMSE1} shows the MSE of the clock skew estimations in nodes $19$ and $5$  as a function of
updating time $\{0, 1, 2,\ldots\}$ for the topology of WSN  shown  in Fig. \ref{randomWSN}.
The number of time-stamp exchange {{rounds} is $N=20$ at the beginning.
Synchronous schedule, asynchronous schedule  and centralized CRB are plotted for comparison.
The synchronous algorithm can only be updated when each node has successfully received updated messages from all its neighboring nodes.
It can be seen from the figure that for both synchronous and asynchronous algorithms, MSEs touch the corresponding CRBs,
{{which are supported by Theorem 3.}
However, due to   {{the} random packet losses, their convergence speeds differ.
Even for high probability of successful transmission ($p_{i, j}=0.99$), the network with synchronous schedule has to  wait for all nodes to receive newly updated
information from all neighbours,
thus {{it presents} slow convergence.
For the same $p_{i,j}$, asynchronous scheduling shows {{extremely} fast convergence, since
each node updates independently.
Furthermore, even with very low probability of successful transmission ($p_{i, j}=0.2$),
asynchronous scheduling can also converge within  $10$ {{iterations}.
However, with such a small $p_{i, j}$, synchronous scheduling would waste most of its time in waiting for updated messages,
and shows extremely slow convergence.
The convergence properties of {{nodes} $5$ and $19$ are also compared in Fig. \ref{25NodeMSE1}.
As node $5$ being a neighbour of the reference node, while node $19$ being much far away,
node $5$ converges faster than node $19$.
Besides, we observe that the further away from the reference node, the larger is the corresponding CRB, i.e.,
$\textrm {CRB}(\alpha_{19})>\textrm {CRB}(\alpha_5)$.
Fig. \ref{25NodeMSE0} shows the corresponding results for the clock offset {{estimation.}
It can be seen from the figure that same conclusions as in Fig. \ref{25NodeMSE1} can be drawn.


Finally, Fig. \ref{MSEvsN} shows the MSE for clock skews and offsets averaged over all nodes versus the number of time-stamp exchange {{rounds} $N$.
$p_{i,j}=0.2$ is assumed for the network.
The MSE is computed after the asynchronous BP algorithm runs for $30$ updating {{iterations}.
$5000$ random network topologies were generated for averaging.
As shown in the figure, the network MSE achieves the best performance as it {{reaches}
the CRB.
This figure also shows that the proposed algorithm can achieve the best performance even under a small number of time-exchange {{rounds}.

\section{Conclusions} \label{Section 6}
In this paper, an asynchronous fully distributed clock skew and offset estimation algorithm for WSNs was proposed.
The algorithm is based on asynchronous BP and is easy to be implemented by exchanging limited information between neighboring sensor nodes.
The proposed algorithm can handle random packet losses and allows some nodes to compute faster and execute more iterations than others.
It was shown analytically that the totally asynchronous algorithm converges regardless of the network topology, and
the MSE of the clock parameter estimates reaches the centralized CRB asymptotically.
Simulations further showed that the asynchronous algorithm converges faster than its synchronous counterpart.
\appendix
\section{Derivative of CRB for Centralized Estimation}  \label{CRB}
We derive the centralized CRB under the assumption that all information over the network
can be gathered in a center.
First, rewrite (\ref{twoequ1}) and (\ref{twoequ2}) as
\begin{equation} \label{twoequ3}
\left[\begin{array}{cc}
c_j(t^2_n)& -1
\end{array} \right]
\underbrace{\left[\begin{array}{c}
\frac{1}{\alpha_j}\\ \frac{\theta_j}{\alpha_j}
\end{array} \right]}_{\bm\beta_j}
= \left[\begin{array}{cc}
c_i(t^1_n)& -1
\end{array} \right]
\underbrace{\left[\begin{array}{c}
\frac{1}{\alpha_i}\\ \frac{\theta_i}{\alpha_i}
\end{array} \right]}_{\bm\beta_i}
 + d_{i,j} + w_{j,n},
\end{equation}
and
\begin{equation} \label{twoequ4}
\left[\begin{array}{cc}
c_j(t^3_n)& -1
\end{array} \right]
\underbrace{\left[\begin{array}{c}
\frac{1}{\alpha_j}\\ \frac{\theta_j}{\alpha_j}
\end{array} \right]}_{\bm\beta_j}
= \left[\begin{array}{cc}
c_i(t^4_n)& -1
\end{array} \right]
\underbrace{\left[\begin{array}{c}
\frac{1}{\alpha_i}\\ \frac{\theta_i}{\alpha_i}
\end{array} \right]}_{\bm\beta_i}
 - d_{j,i} - w_{i,n}.
\end{equation}
Stacking (\ref{twoequ3}) and (\ref{twoequ4}) in matrix form with the assumption $d_{i,j}=d_{j,i}$, we have
\begin{equation} \label{twoequ5}
\underbrace{\left[\begin{array}{cc}
c_j(t^2_1)& -1 \\
\vdots & \vdots\\
c_j(t^2_N)& -1 \\
c_j(t^3_1)& -1 \\
\vdots & \vdots\\
c_j(t^3_N)& -1 \\
\end{array} \right]}_{\triangleq\bm T_{j,i}}
\bm\beta_j
-
\underbrace{\left[\begin{array}{cc}
c_i(t^1_1)& -1 \\
\vdots & \vdots\\
c_i(t^1_N)& -1 \\
c_i(t^4_1)& -1 \\
\vdots & \vdots\\
c_i(t^4_N)& -1 \\
\end{array} \right]}_{\triangleq\bm T_{j,i}}
\bm\beta_i
-
d_{i,j}
\left[\begin{array}{c}
\bm 1_N \\
-\bm 1_N\\
\end{array} \right]
=
\underbrace{\left[\begin{array}{c}
w_{j,1 }  \\
\vdots  \\
w_{j,N} \\
-w_{i,1 }  \\
\vdots  \\
-w_{i,N } \\
\end{array} \right]}_{\triangleq\bm n_{j,i}}
\end{equation}
where $\bm 1_N$ is an all one $N$ dimensional vector and
$\bm n_{j,i}\sim \N(\bm n_{j,i}|\bm 0, \textrm{diag}[\sigma^2_j, \sigma^2_i]\otimes \bm I_N)$ {{where}
the symbol $\otimes$  denotes the Kronecker product.

Define $\bm y\in \mathbb{{R}}^{2N{|\mathcal{E}|}\times 1}$ with $-\bm T_{1,i}\bm \beta_{1}$ arranged in ascending
{{order with respect to index $i$}, with
$i\in \mathcal{I}(1)$
and the remaining elements being zeros, and define
$\bm \xi\triangleq[{\bm \beta}^T_2,\ldots, \bm \beta^T_M, \bm d^T]^T$
with vector $\bm d$ containing elements $d_{i,j}$ {{with ascending order first with respect to $j$ and then with respect to $i$}.
Then stacking (\ref{twoequ5}) for all $i$ and $j$, we obtain
\begin{equation}\label{blocklinear}
\bm y= \bm H\bm \xi + \bm n,
\end{equation}
where
$\bm n$ contains $\bm n_{i,j}$ {{with ascending order first with respect to $j$ and then with respect to $i$}.
Notice that $\bm n \sim \N(\bm n|\bm0,\bm \Delta)$ with
$\bm \Delta$ is a block diagonal matrix containing $\bm \Delta_{i,j}=\textrm{diag}[\sigma^2_j, \sigma^2_i]\otimes \bm I_N$ as diagonal block.
Since (\ref{blocklinear}) is a standard linear model, the CRB for $\bm \xi$ is given by $\textrm{CRB}(\bm \xi)=\big[\bm H^T\bm \Delta^{-1}\bm H\big]^{-1}$ \cite{Kay}.

The ultimate goal is to estimate the clock offsets and skews $\bm\zeta \triangleq [\theta_{2}, \alpha_{2}, \ldots, \theta_{M}, \alpha_{M} ]^T$.
Since $\bm \xi$ is a related to $\bm\kappa\triangleq[\bm\zeta^T, \bm d^T]^T$ through a transformation, thus we
{{can express the CRB} matrix of $\bm\zeta$ as \cite{Kay}
\begin{equation}\label{CRBreal}
\mbox{CRB}(\bm\zeta)= \bigg(\frac{\partial \bm\kappa}{\partial \bm\xi}\bigg)
\textrm{CRB}(\bm \xi)
\bigg(\frac{\partial \bm\kappa}{\partial \bm\xi} \bigg)^T.
\end{equation}
It can be {{easily inferred } that
$\partial \bm\kappa/\partial \bm \xi =
\left[\begin{array}{cc}
\bm\Sigma &  \bm 0 \\
\bm 0&   \bm I_{\frac{1}{2}|\mathcal{E}|}]
\end{array} \right]
$ with
$\bm\Sigma$ being {{a} $2(M-1)$-by-$2(M-1)$
block diagonal matrix with
the $m^{th}$ diagonal block being
$
\left[\begin{array}{cc}
-\alpha_{m+1}\theta_{m+1} & \alpha_{m+1} \\
 -\alpha^2_{m+1}& 0\\
\end{array} \right]
$.

\begin{figure}[!ht]
\centering
\epsfig{file=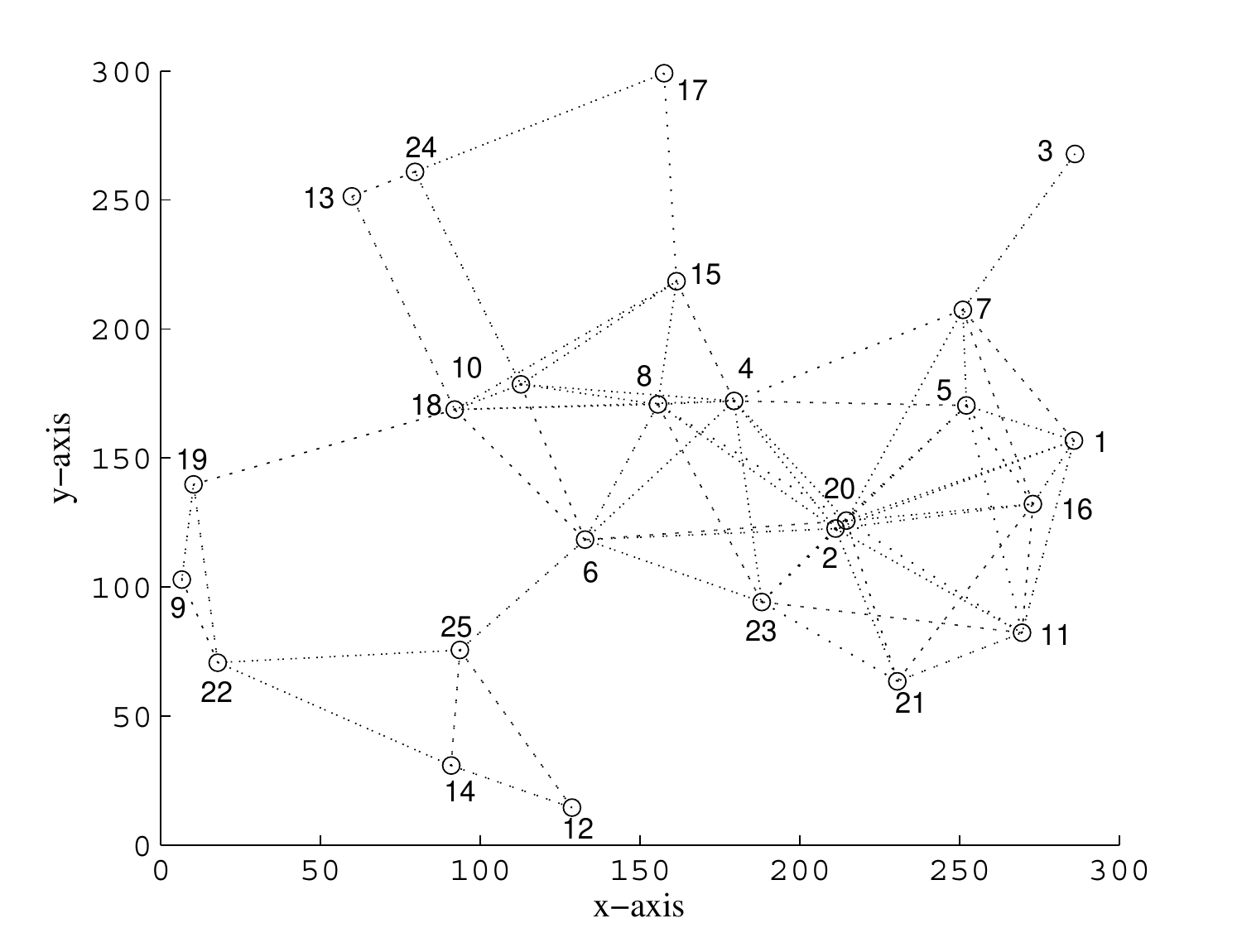, width=4in}
\caption{WSN topology with $25$ nodes randomly distributed.}
\label{randomWSN}
\end{figure}

\begin{figure}[!ht]
\centering
\epsfig{file=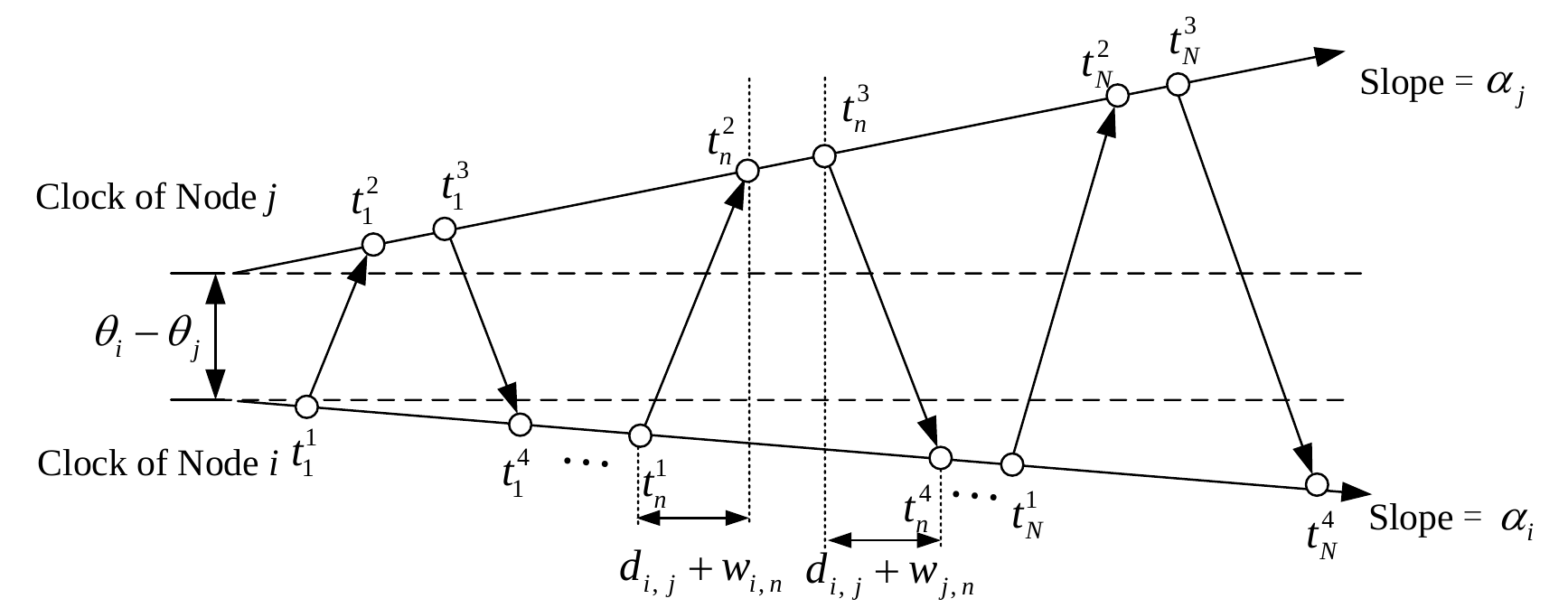, width=4in}
\caption{Two way message exchange between node $i$ and $j$ in the WSN. }
\label{TwoWay}
\end{figure}

\begin{figure}[!ht]
\centering
\epsfig{file=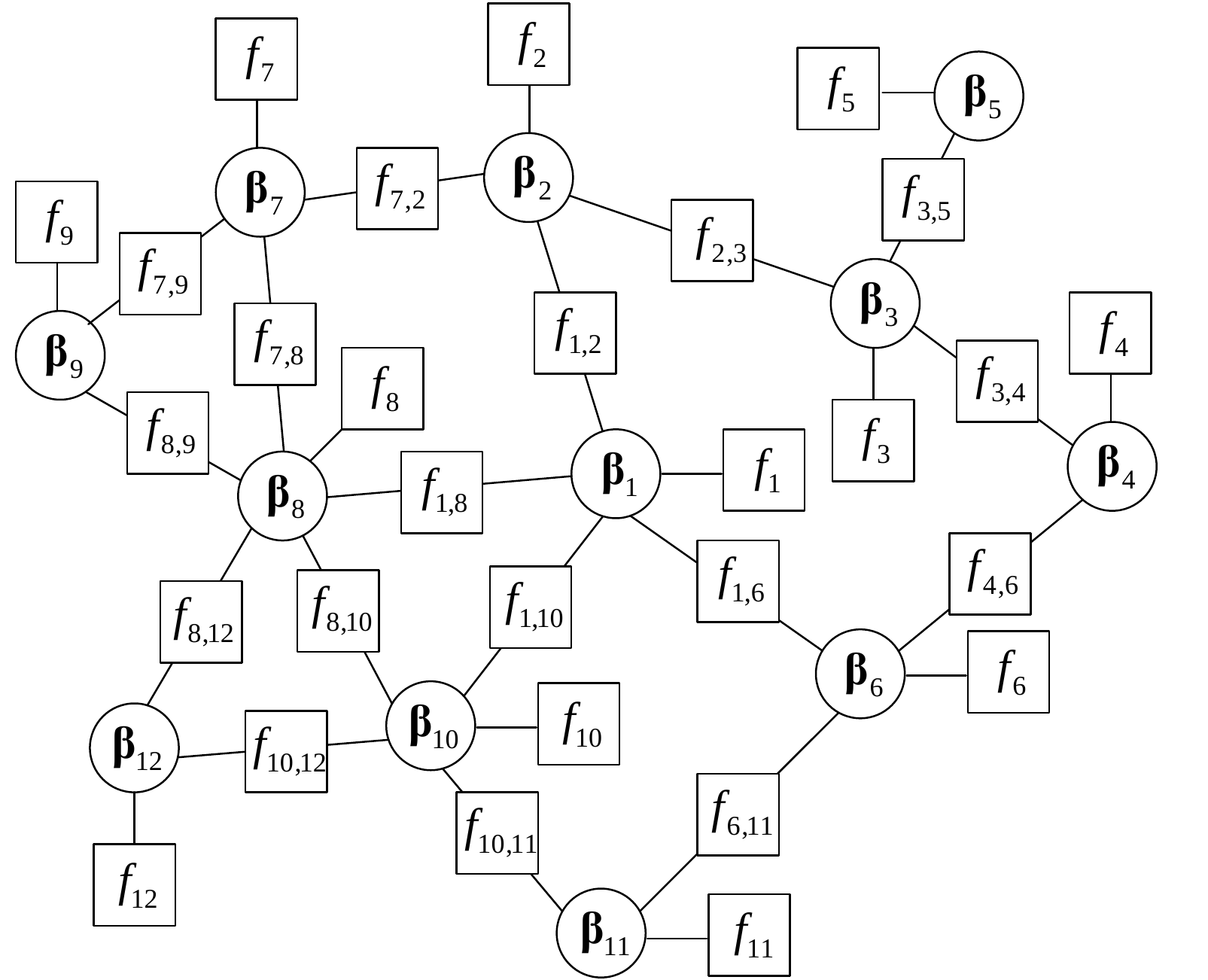, width=3.5in}
\caption{The factor graph for clock synchronization in a WSNs.}
\label{FG}
\end{figure}

\begin{figure}[!ht]
\centering
\epsfig{file=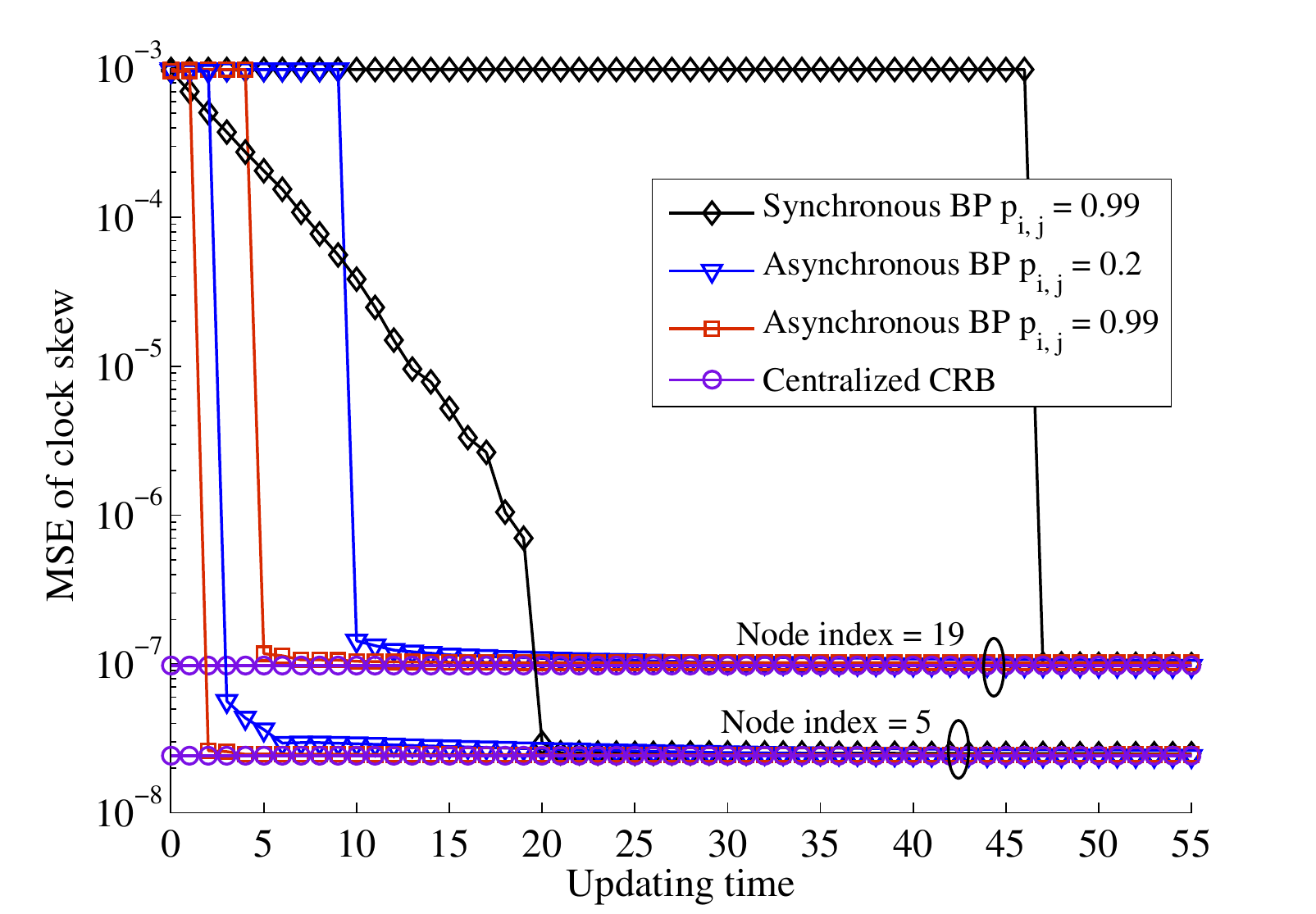, width=4in}
\caption{Convergence performance of estimated clock skew at two nodes.}
\label{25NodeMSE1}
\end{figure}

\begin{figure}[!ht]
\centering
\epsfig{file=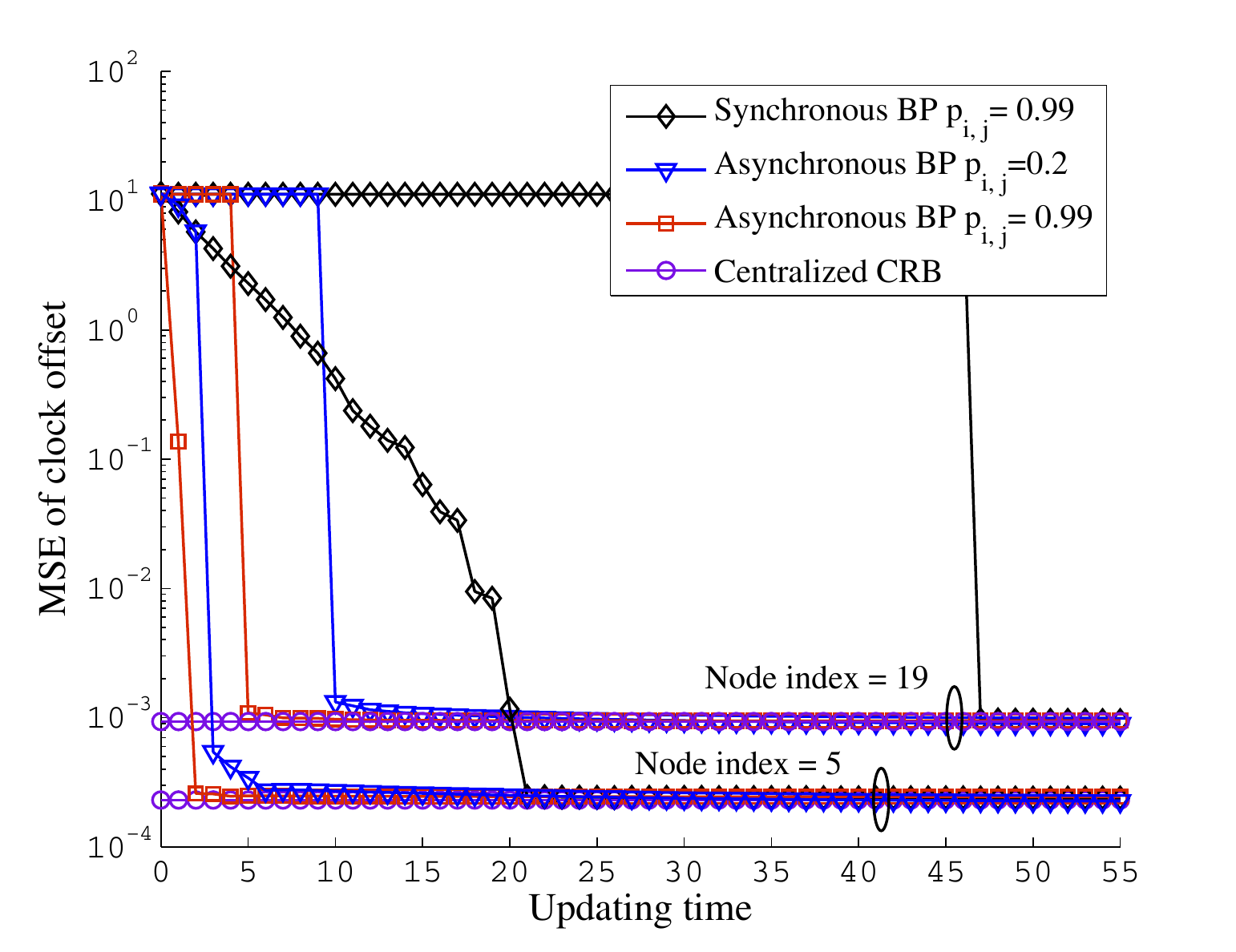, width=4in}
\caption{Convergence performance of estimated clock offset at two nodes.}
\label{25NodeMSE0}
\end{figure}

\begin{figure}[!ht]
\centering
\epsfig{file=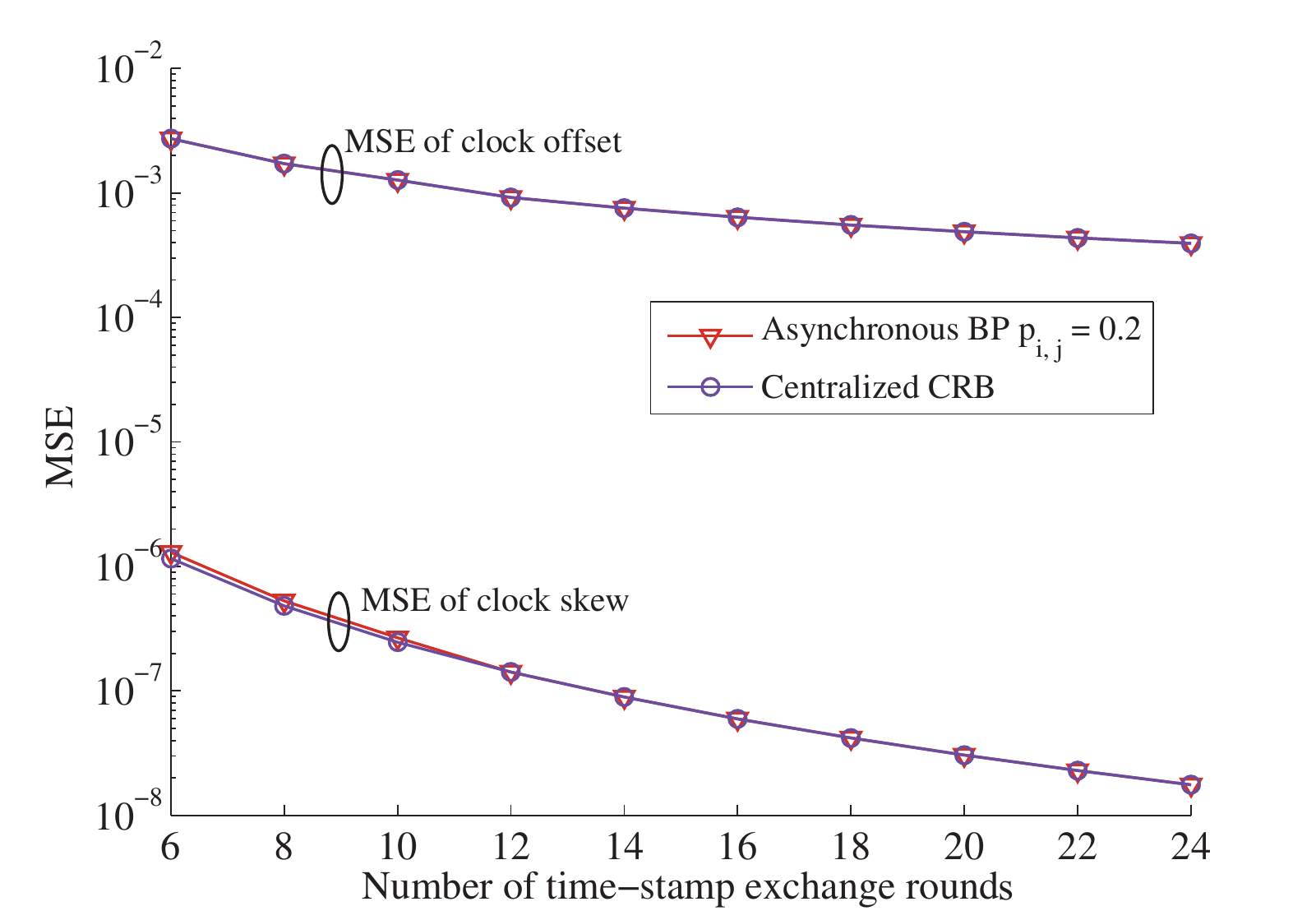, width=4in}
\caption{MSE of clock skew and offset averaged over the whole network under random network topologies.}
\label{MSEvsN}
\end{figure}

\end{document}